# Interlayer Coupling Driven Correlated and Charge-Ordered Electronic States in a Transition Metal Dichalcogenide Superlattice

Yiwei Li[1,*,†], Lixuan Xu[2,*], Shihao Zhang[3,*], Lanxin Liu[4,*], Yifan Zhou[1], Qiang Wan[1], Shiwei Chen[2], Shiheng Liang[2], Yulin Chen[5,6,7], Yi-feng Yang[8], Xuan Luo[4,‡], Yuping Sun[4,9,10], Nan Xu[1,11,§], and Zhongkai Liu[5,6,¶]

[1]*Institute for Advanced Studies (IAS), Wuhan University, Wuhan 430072, China*

[2]*Department of Physics, Hubei University, Wuhan 430062, China*

[3]*School of Physics and Electronics, Hunan University, Changsha 410082, China*

[4]*Key Laboratory of Materials Physics, Institute of Solid State Physics, HFIPS, Chinese Academy of Sciences, Hefei 230031, China*

[5]*School of Physical Science and Technology, ShanghaiTech University, Shanghai 201210, China*

[6]*ShanghaiTech Laboratory for Topological Physics, Shanghai 201210, China.*

[7]*Department of Physics, University of Oxford, Oxford OX1 3PU, UK*

[8]*Beijing National Laboratory for Condensed Matter Physics, Institute of Physics, Chinese Academy of Science, Beijing 100190, China*

[9]*Anhui Province Key Laboratory of Condensed Matter Physics at Extreme Conditions, High Magnetic Field Laboratory, HFIPS, Chinese Academy of Sciences, Hefei, 230031, China*

[10]*Collaborative Innovation Center of Microstructures, Nanjing University, Nanjing 210093, China*

[11]*Wuhan Institute of Quantum Technology, Wuhan 430026, China*

*These authors contributed equally to this work.

†Contact author: yiweili@whu.edu.cn

‡Contact author: xluo@issp.ac.cn

§Contact author: nxu@whu.edu.cn

¶Contact author: liuzhk@shanghaitech.edu.cn

## Abstract

**$4Hb\text{-}TaS_2$, a van der Waals superlattice comprising alternate stacked Ising superconducting $1H\text{-}TaS_2$ and cluster Mott insulating $1T\text{-}TaS_2$, exhibits emergent properties beyond those of its constituent layers. Notable phenomena include time-reversal-symmetry-breaking superconductivity and spontaneous vortex phases, which are driven by nontrivial interlayer interactions that remain debated. Using area-selective angle-resolved photoemission spectroscopy, we provide direct spectroscopic evidence of such interaction by systematically probing the electronic structures of 1T- and 1H-terminted surfaces of $4Hb\text{-}TaS_2$. The metallic states of subsurface 1H-layers are folded to the Brillouin zone center by the $\sqrt{13} \times \sqrt{13}$ modulation of the surface 1T-layer, forming chiral "windmill" Fermi surfaces via Umklapp scattering. These conducting states further hybridize with the incipient flat band of the surface 1T-layer, producing a Kondo-like peak at the Fermi level. Interlayer charge transfer induces distinct 3×3 and 2×2 charge orders on the surface and subsurface 1H-layers, respectively, which result in characteristic segmented Fermi surfaces and dichotomously shift the van Hove singularities. These findings reconcile the competing Kondo and Mott-Hubbard models in this material and emphasize the interplay of flat bands, van hove singularities, charge orders, and unconventional superconductivity in correlated superlattices.**

# I. INTRODUCTION

Interlayer interactions are central to van der Waals (vdW) layered materials, enabling a wealth of exotic emerging states in artificial vdW heterostructures [1,2]. In contrast to artificial stackings of atomically thin layers, where twist angles and lattice mismatch are key design parameters, natural vdW materials form stable bulk crystals from two-dimensional (2D) layers with nearly identical lattice constants [3]. Free from sophisticated nanoscale fabrications, they offer macroscopic homogeneity, broad experimental accessibility, and strongly coupled systems beyond the tear-and-stack approach. A prominent example is the $MnBi_{2+2n}Te_{4+3n}$ superlattice family, built from magnetic $MnBi_2Te_4$ and nonmagnetic $Bi_2Te_3$ layers, which hosts quantum anomalous Hall insulators, axion insulators, and intrinsic magnetic topological insulators [4-6]. The pivotal role of interlayer coupling is also demonstrated in bismuth halides [7-9] and 2M-phase transition metal dichalcogenides (TMDCs) [10,11], composed of quantum spin Hall layers, where variations in stacking geometry and interaction strength generate diverse topological phases.

4Hb-$TaS_2$ is a natural stable polymorph of bulk TMDC [12] consisting of vdW heterostructures of trigonal prismatic (1H) and octahedral (1T) $TaS_2$ layers [Fig. 1(a)]. Monolayer 1H-$TaS_2$ lacks inversion symmetry and becomes superconducting below about 3 K with Ising pairing [13,14]. Monolayer 1T-$TaS_2$ undergoes a commensurate charge density wave (CDW) transition at about 350 K and forms a Star of David (SoD) cluster of Ta atoms with a $\sqrt{13} \times \sqrt{13}$ superstructure [15]. This state is described as a cluster Mott insulator and proposed to be a quantum spin liquid candidate [15,16].

Previous transport and X-ray diffraction studies have revealed two CDW transitions in 4Hb-$TaS_2$: one at 315 K, associated with the $\sqrt{13} \times \sqrt{13}$ superstructure on 1T-layers, and another at 22 K, arising from the 2×2 charge orders on 1H-layers [17-19]. A superconducting transition occurs at about 3 K [17-19], comparable to that of monolayer 1H-$TaS_2$. Recently, 4Hb-$TaS_2$ has drawn considerable attention owing to its unprecedented properties, including time-reversal-symmetry-breaking (TRSB) in both the superconducting state [20] and a spontaneous vortex state [21], multicomponent order parameters [22,23], topological edge modes [24], and finite momentum superconducting pairing states [25]. Yet, the precise nature of the interplay between 1H- and 1T-layers remains unsettled, with the central controversy revolving around whether the system is described as a Mott-Hubbard model mediated by interlayer charge transfer, or a superconducting Kondo lattice scenario arise from the exchange interaction between localized magnetic moments and conduction electrons [26-29].

In this Article, we comprehensively investigate the electronic structures of 1T- and 1H-terminations of 4Hb-$TaS_2$ and reveal direct evidence of interlayer interaction between adjacent 1T- and 1H-layers using area-selective angle-resolved photoemission spectroscopy (ARPES) [30]. With a low photon energy (<100 eV), the probing depth of ~1 nm [31] spans approximately two vdW layers (the interlayer distance of 4Hb-$TaS_2$ is ~6 Å). In contrast to local probes such as scanning tunneling microscopy (STM) and bulk sensitive transport measurements, ARPES offers an ideal compromise: sufficient surface sensitivity to distinguish electronic structures from different layers, together with a probing depth adequate to capture interlayer interactions. These interactions manifest as Umklapp-scattered metallic states of the 1H-layers under the $\sqrt{13} \times \sqrt{13}$ modulation of the 1T-layer. The spectral intensity enhancement near the Fermi level ($E_F$) reveals the coupling of conducting states of 1H-layers to the incipient flat band (FB) of 1T-layer SoDs. Furthermore, CDW states and van Hove singularities (VHS) of surface and subsurface 1H-layers are mediated by interlayer charge transfer, leading to distinct segmented Fermi surfaces. Our findings establish a benchmark for theoretical modelling of the exotic properties in 4Hb-$TaS_2$ and offer insights into the strongly correlated electrons in vdW superlattices.

## II. RESULTS

### A. Termination-dependent electronic structures

4Hb-$TaS_2$ crystallizes in the hexagonal space group $P6_3/mmc$ (No. 194) with the lattice constants $a$ = 3.332 Å and $c$ = 23.62 Å [12]. Each unit cell comprises four alternate stacked 1T- and 1H-$TaS_2$ layers along the $c$ axis [Fig. 1(a)]. The Ta atoms within each layer are aligned with the $c$ axis, whereas the next neighboring layers are related by a 180° in-plane rotation. Cleaving the bulk crystals naturally exposes either 1T- or 1H-terminations [Fig. 1(a)].

With a photon beam size of ~1×1 $\mu m^2$, these two terminations are distinguished by their characteristic core level spectra [Figs. 1(b) and 1(c)]. For both terminations, three peaks are resolved within each of the Ta $4f_{7/2}$ and $4f_{5/2}$ submanifolds, but with noticeable energy shifts and sharp intensity contrasts. The low-binding-energy peaks primarily originate from Ta atoms on the 1H-layer, with a minor contribution from the center Ta sites ($Ta_a$) of the SoD on the 1T-layer. The middle-binding-energy and high-binding-energy peaks are attributed to the middle ($Ta_b$) and outer Ta sites ($Ta_c$) on the 1T-layer, respectively, and are much more pronounced on the 1T-termination than on the 1H-termination. Due to work function differences [32], charge transfer from the 1T-layer to the 1H-layer leads to polar surfaces [see Fig. 1(a)]. Since the surface 1T-layer is less hole-doped than the subsurface 1T-layer, its core-level peaks appear at higher binding energies on the 1T-termination compared to the 1H-termination. In contrast, the 1H-layer derived core levels remain largely unaffected by charge transfer due to the good metallicity of the 1H-layers. These results are consistent with previous X-ray photoemission spectroscopy studies [33-36].

Termination-dependent electronic structures are systematically measured by area-selective ARPES at 46 K [Figs. 1(d)-1(g)]. For the 1H-termination [Fig. 1(d)], the Fermi surface is dominated by metallic states from the surface 1H-layer. The $E_F$ lies near a critical point where the shoulders of the “dogbone” electron pockets merge, giving rise to a VHS at about 0.58 ΓK, or (0.19, -0.19, 0) in reciprocal lattice unit [Figs. 1(d) and 1(f)]. The corresponding result of the 1T-termination shows double-walled hexagonal hole pockets around Γ and circular hole pockets around K, originating from the subsurface 1H-layer [Fig. 1(e)]. Compared with the surface 1H-layer, the subsurface 1H-layer is more electron-doped as it is sandwiched by electron-donating 1T-layers [see Fig. 1(a)], which drives the VHS below $E_F$ [Figs. 1(e) and 1(g)]. Additionally, the striking chiral “windmill” Fermi surfaces observed around Γ [Fig. 1(e)], previously misattributed to the surface 1T-layer [20,25], are argued in the following section to originate instead from Umklapp-scattered 1H-layer states under the $\sqrt{13} \times \sqrt{13}$ modulation.

The band dispersion of the 1H-termination reveals intense metallic bands from the surface 1H-layer, accompanied by much weaker features from the insulating subsurface 1T-layer [Fig. 1(f)]. The valence band maximum lies ~80 meV below the Fermi level. In contrast, the 1T-termination exhibits two sets of 1T-layer-derived bands [Fig. 1(g)], with valence band maxima at ~280 meV and ~80 meV. The lower band with stronger intensity is attributed to the less hole-doped surface 1T-layer, while the upper band shows a doping level comparable to that of the 1H-termination, suggesting a bulk 1T-layer origin. Nevertheless, additional investigations are required to exclude the possibility of intrinsic surface band splitting arising from unknown mechanisms.

These band assignments are corroborated by theory and experiment. The 1H-layer bands match calculated 1H-monolayer dispersions after a Fermi-level adjustment (Fig. S1 [37]), while the 1T-layer bands follow the $\sqrt{13} \times \sqrt{13}$ mini−Brillouin zone periodicity (Fig. S2 [37]), also consistent with the upshifted 1T-monolayer bands [15]. The persistent degeneracy of the 1H-layer band along ΓM measured with varying photon energies (Fig. S3 [37]), further confirms the 2D nature of 4Hb-$TaS_2$ with negligible interaction between next-neighboring layers.

## B. Interaction between neighboring 1T- and 1H-layers

Figure 2 illustrates the non-interacting electronic structures of the 1T-layer as an electron donor in the 1T/1H heterostructures. In a free-standing 1T-monolayer hosting the $\sqrt{13} \times \sqrt{13}$ SoD superstructure, one unpaired electron localizes at the central Ta atom of each SoD and forms a half-filled, narrow FB [Figs. 2(a), 2(d), and 2(g)]. It evolves into a Mott insulator when electron correlations are included (Fig. S4 [37]). By comparing the doping levels of surface and subsurface 1H-layers, the interlayer charge transfer is estimated at 0.0715±0.0025 electron per Ta atom, or 0.92±0.04 electron per SoD, based on a rigid band model, consistent with the previous study [38] and the ab initio calculation (Fig. S5 [37]). Consequently, the filling factor on the surface 1T-layer in 4Hb-$TaS_2$ is 0.08±0.04 electron per SoD, forming an incipient FB that lie just above $E_F$ [Figs. 2(b), 2(e), and 2(h)], consistent with STM observations of ~12% electron-filled and ~88% electron-empty SoDs [28,29]. In contrast, the subsurface 1T-layer donates electrons to both sides of the 1H-layers, resulting in electron-void SoDs and remote empty FB away from $E_F$ [Figs. 2(c), 2(f), and 2(i)].

High resolution Fermi surface mapping of the 1H-termination near Γ agrees with the electronic structures of the surface 1H-layer [Figs. 3(a) and 3(b)], showing no evident signatures of modulation from the subsurface 1T-layer. Theoretically predicted replicas of 1H-layer derived bands, scattered by the $\sqrt{13} \times \sqrt{13}$ superlattice vectors, carry infinitesimal spectral weight [Fig. 3(c)] (more details can be found in Fig. S6 [37]). This suppression is consistent with the weak coupling between the metallic states and the remote empty FB, which is also supported by the integrated energy-distribution-curve (EDC) showing an unperturbed Fermi-Dirac line profile [Fig. 3(c)].

In addition to the subsurface 1H-layer derived Fermi surfaces, we identify the striking chiral "windmill" features on the 1T-termination [Fig. 3(e)] as Umklapp scattered metallic states arising from the $\sqrt{13} \times \sqrt{13}$ modulation, as illustrated in the simulation [Fig. 3(f)]. The dispersions of the scattered bands replicate those of the original bands, as evidenced by spectral cuts linked by the $\sqrt{13} \times \sqrt{13}$ superstructure reciprocal vectors [Fig. 3(g)]. A pronounced spectral weight enhancement near $E_F$ is observed, evidence of the metallic states hybridized with the incipient FB [Fig. 3(h)] (more details can be found in Fig. S7 [37]). The integrated EDC exhibits a Kondo-like peak at $E_F$ and it decreases in intensity with increasing temperatures, with a fitted half width at half maximum of 6.5 meV at 16 K (Fig. S8 [37]), consistent with STM measurement at the center of electron-filled SoDs [28,29]. The spectral weight modulation of the "windmill" feature might be attributed to the weak dispersion of the incipient FB.

### C. Charge-ordered electronic states of the 1H-layers

The surface and subsurface 1H-layers host distinct CDW states driven by interlayer charge transfer (Fig. S9 [37]). On the surface 1H-layer, six $Ta_b$ atoms shift toward the central $Ta_a$ atom to form a hexagonal seven-atom cluster, while $Ta_c$ atoms remain fixed, giving rise to a 3×3 CDW observed by STM on the 1H-termination of 4Hb-$TaS_2$ [39] and previously reported in bulk 2H-$TaS_2$ and 2H-$TaSe_2$ [40,41]. On the subsurface and bulk 1H-layers, a 2×2 CDW emerges as four $Ta_b$ atoms distort toward the central $Ta_a$ atom, forming triangular four-atom clusters, as revealed by bulk-sensitive X-ray diffraction [19]. A doping-driven 3×3 to 2×2 CDW transition has also been reported in alkali-metal-doped 2H-$TaS_2$ films [42] and 2H-$TaSe_2$ crystals [43].

Intriguingly, these interlayer charge transfer mediated charge orders significantly alter the low-energy electronic structures, as shown in Fig. 4. The inner K-centered circular pocket evolves into discrete arc features under CDW modulations, as revealed by Fermi surface measurements at 46 K and 16 K [Fig. 4(a)] (more details can be found in Fig. S10 [37]) and reproduced by the ab initio calculations [Fig. 4(b)]. On the surface 1H-layer, the spectral weight near the KΓ ($\theta = 0$) and KM ($\theta = 60°$) directions is strongly suppressed [Fig. 4(g)], resulting in three separated arcs [Fig. 4(f)]. In contrast, the Fermi surface of the subsurface 1H-layer develops a nine-arc pattern [Fig. 4(f)], with pronounced intensity peaks near the KM ($\theta = 60°$) directions and $\theta \approx 18°$, but suppressed intensity near the KΓ ($\theta = 0$) directions and $\theta \approx 36°$ [Fig. 4(h)]. The gap-like features on the reconstructed Fermi surfaces align with the CDW vectors (Fig. S9 [37]), consistent with hybridization between CDW-folded and original bands in the Fermi nesting model [41].

The saddle dispersions near the VHS are resolved by cuts along perpendicular momentum directions, which exhibit opposite curvatures [Figs. 4(c) and 4(d)]. Notably, the 3 ×3 and 2×2 CDWs shift the VHS in opposite directions: on the surface 1H-layer, it is raised from $E_{\mathrm{VHS}} \approx E_{\mathrm{F}} - 10$ meV to $E_{\mathrm{VHS}} \approx E_{\mathrm{F}} + 10$ meV, while on the subsurface 1H-layer, it is lowered from $E_{\mathrm{VHS}} \approx E_{\mathrm{F}} - 40$ meV to $E_{\mathrm{VHS}} \approx E_{\mathrm{F}} - 60$ meV. As $E_{\mathrm{F}}$ crosses the critical $E_{\mathrm{VHS}}$, the topology of the Fermi surfaces undergoes a Lifshitz transition: the M-centered "dogbone" pockets at $E_{\mathrm{F}} < E_{\mathrm{VHS}}$ merge and evolve into the outer Γ-centered and K-centered circular pockets at $E_{\mathrm{F}} > E_{\mathrm{VHS}}$ [Figs. 4(a), 4(b), and 4(e)].

## III. DISCUSSION

Our experimental evidence of the hybridization between the conducting states of the subsurface 1H-layer and the incipient FB of the surface 1T-layer [Figs. 5(a) and 5(b)] resolve the debate on the interlayer interaction [26-29]. The pronounced interlayer charge transfer results in a low electron filling of the incipient FB (~0.1), driving the system away from the Mott-Hubbard region and towards the non-interacting limit. However, this hybridization can also induce exotic physical consequences. The presence of multiband superconductivity, with coexisting dispersive bands and incipient FB, could favor complex cross-orbital pairings, potentially leading to multicomponent and topological nodal-point superconductivity [22-24]. With the established interlayer band hybridization, one would naturally expect the appearance of a Kondo resonance peak at electron-filled SoD centers, as observed by STM on 1T/1H bilayers [26], in a scenario where the superconducting states are coupled to dilute magnetic impurities [44]. Furthermore, incipient FB may enhance superconductivity and stimulate heavy-fermion behaviors [45-47].

In contrast to the 1T-terminated surface, TRSB in the bulk superconductivity of 4Hb-$TaS_2$, evidenced by enhanced muon-spin relaxation below the superconducting transition temperature [20], may originate solely from 1H-layers. Due to interlayer charge transfer, the bulk 1T-layers are over hole-doped Mott insulators, with zero DOS at $E_F$ [Fig. 5(a)], while the bulk 1H-layers provide all the superconducting states near the 2D VHS [Fig. 5(c)]. Chiral *p*-wave superconducting pairings on a hexagonal lattice can naturally emerge from these VHS at non-time-reversal invariant momenta [48], which is consistent with the observed nodeless bulk superconductivity [49,50]. Furthermore, signatures of TRSB have recently reported in 2H-TMDC films [51,52]. The lack of interlayer band hybridization in the bulk [see Fig. 3(d)] suggests that 1T-layers primarily function as intercalated layers that decouples 1H-layers. An alternative scenario is that the mirror-symmetry-breaking $\sqrt{13} \times \sqrt{13}$-superstructure on the 1T-layer could induce unconventional superconductivity through a proximity effect [25], reminiscent of the case in the chiral molecules intercalated 2H-$TaS_2$ [53].

Finally, TMDC materials crystallize in rich polymorphs and are versatile with intercalations [54,55]. Bulk vdW superlattices comprising TMDC layers thus provide a fertile playground for exploring novel quantum phases arise from interlayer coupling. Area-selective ARPES measurements with state-of-the-art spatial resolution provide a comprehensive study of low-energy electronic structures of 4Hb-$TaS_2$ and a benchmark for further theoretical modeling. This approach can be readily extended to complex vdW TMDC superlattices [56-58].

## IV. CONCLUSION

In conclusion, we have employed advanced area-selective ARPES to provide critical insights into the emergent phenomena in the TMDC superlattice material 4Hb-$TaS_2$. Our findings fundamentally revise the understanding of its electronic structure. We establish that the interlayer-modulated 1H-bands, previously misidentified as coherent 1T-layer states, are scattered by the $\sqrt{13} \times \sqrt{13}$-superstructure of the neighboring 1T-layer. Furthermore, we identify a Kondo-like peak at the Fermi level, which we attribute to the interlayer hybridization of metallic 1H-bands with the incipient FB of the surface 1T-layer. This profound interlayer coupling also mediates charge orders on different 1H-layers, leading to the segmented Fermi surfaces and dichotomous energy shift of the VHS. These results underscore the determinative role of interlayer interactions in driving the correlated electronic states and unconventional superconductivity in 4Hb-$TaS_2$.

## V. METHODS

### A. Sample synthesis

4Hb-$TaS_2$ single crystals were grown by the chemical vapor transport method with iodine as the transport agent. Ta and S powders with a mole ratio of 1:2 were weighted and mixed with 0.15 g of $I_2$, which were placed into silicon quartz tubes. These tubes were sealed under high vacuum and heated for 10 days in a two-zone furnace, where the temperature of the source and growth zones were fixed at 780 °C and 680 °C, respectively. Then the quartz tubes were removed from the furnace and quenched in the ice water mixture [18].

### B. ARPES measurements

Area-selective ARPES measurements were performed at beamline BL07U, Shanghai Synchrotron Radiation Facility, China, with a beam spot size of 1×1 $\mu m^2$. The samples were cleaved in situ under ultra-high vacuum below $5\times10^{-11}$ Torr. Data were collected by Scienta DA30L analyzer. The optimized energy resolutions were below 20 meV, and the angle resolution was 0.1°, respectively [59].

### C. Electronic band structure calculations

We performed first-principles calculations using the Vienna Ab-initio Simulation Package (VASP) based on density functional theory [60,61]. Throughout all our calculations, we employed the Perdew-Becke-Ernzerhof functional within the generalized gradient approximation [62] to describe the exchange-correlated potential. The plane wave cutoff energy was set to 400 eV. For the calculations concerning the Mott phase in the $\sqrt{13}\times\sqrt{13}$-modulated 1T-$TaS_2$ monolayer, we used a Hubbard parameter U = 2 eV for the *d* orbitals of the Ta atom at the center of the SoD. In the calculations of the Fermi surface of the 2×2 and 3

×3 charge-ordered 1H-$TaS_2$ monolayers, we unfolded the energy bands to the Brillouin zone of the primitive cell, applying a Gauss smearing factor of 30 meV. For the slab model calculations, we utilized a model consisting of four unit cells (16 vdW layers) of 4Hb-$TaS_2$ to account for interlayer charge transfer in both the bulk and surface regions.

## ACKNOWLEDGEMENTS

Y.L. acknowledges support from the Natural Science Foundation of Hubei Province, China (Grant No. 2024AFB935). L.X. acknowledges support from the National Natural Science Foundation of China (Grants No. 12404075) and Natural Science Foundation of Wuhan (Grant Grants No. 202504061020153). S.Z. was supported by the National Natural Science Foundation of China (No. 12304217), the National Key Research and Development Program of China (No. 2024YFA1410300), the Natural Science Foundation of Hunan Province (No. 2025JJ60002) and the Fundamental Research Funds for the Central Universities from China (No. 531119200247). X.L. and Y.S. acknowledge support from the National Key Research and Development Program of China (Grants No. 2021YFA1600201and 2023YFA1607402). Y.S. acknowledges support from the National Nature Science Foundation of China (Grants No. 12274412) and Systematic Fundamental Research Program Leveraging Major Scientific and Technological Infrastructure, Chinese Academy of Sciences under Contract No. JZHKYPT-437 2021-08. Y.Y. acknowledge support from by the National Natural Science Foundation of China (No. 12474136) and the National Key Research and Development Program of China (2022YFA1402203).

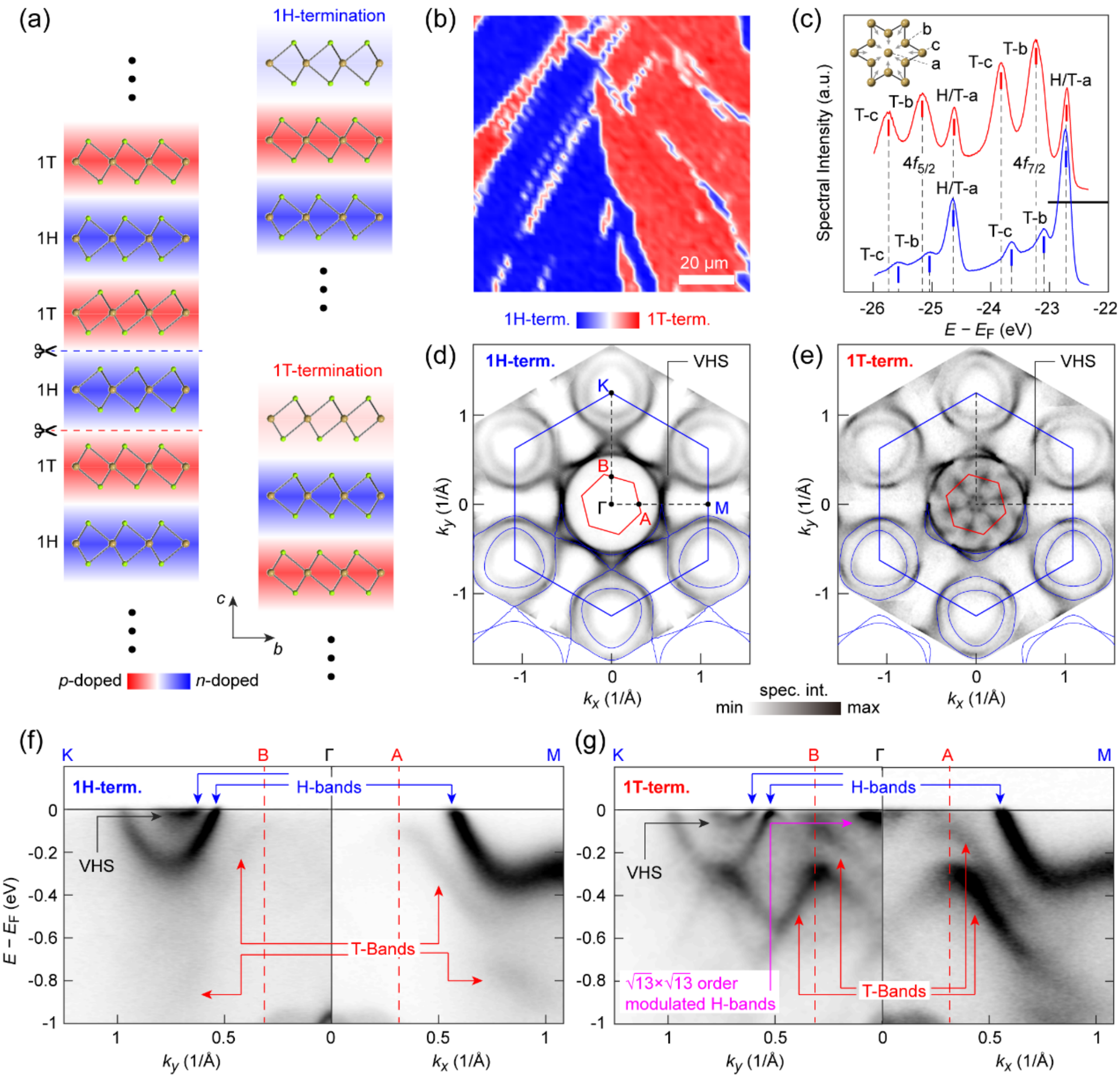

FIG. 1. Termination-dependent electronic structures of 4Hb-$TaS_2$. (a) Schematic illustration of the crystal structure of 4Hb-$TaS_2$ consisting of alternate stacked 1H- and 1T-$TaS_2$, exposing 1H- and 1T-terminations. The interlayer charge transfer results in hole-doped 1T-layers (red shaded areas) and electron-doped 1H-layers (blue shaded areas). (b) Photoemission spatial scanning showing 1H-termination regions (blue) and 1T-termination regions (red). c, Core level spectra of Ta $4f_{5/2}$ and $4f_{7/2}$ measured on the 1T- and 1H-terminations, as indicated by the red and blue curves, respectively. Inset: Illustration of the thirteen-atom-cluster SoD with three types of Ta atoms under different chemical environments. T-a,b,c indicates the core level peaks derived from $Ta_{a,b,c}$ atoms on the 1T-layers. H indicates the core level peaks derived from the 1H-layers. (d),(e) Fermi surface mapping measurements on the 1H-termination (d) and 1T-termination (e), appended with the 2D Brillouin zone (BZ) of the 1H-layer (the blue hexagon), the mini-BZ of the $\sqrt{13} \times \sqrt{13}$ superstructure of the 1T-layer (the red hexagon), and calculated Fermi surfaces of monolayer 1H-layers with adjusted $E_F$ (blue curves). (f),(g) Band dispersion measurements on the 1H-termination (f) and 1T-termination (g), respectively, along high-symmetry momentum directions as indicated by the dashed lines in (c) and (d). Bands derived from 1H-layers, 1T-layers, and scattered states due to 1T/1H interlayer coupling are marked by blue, red, and magenta arrows, respectively.

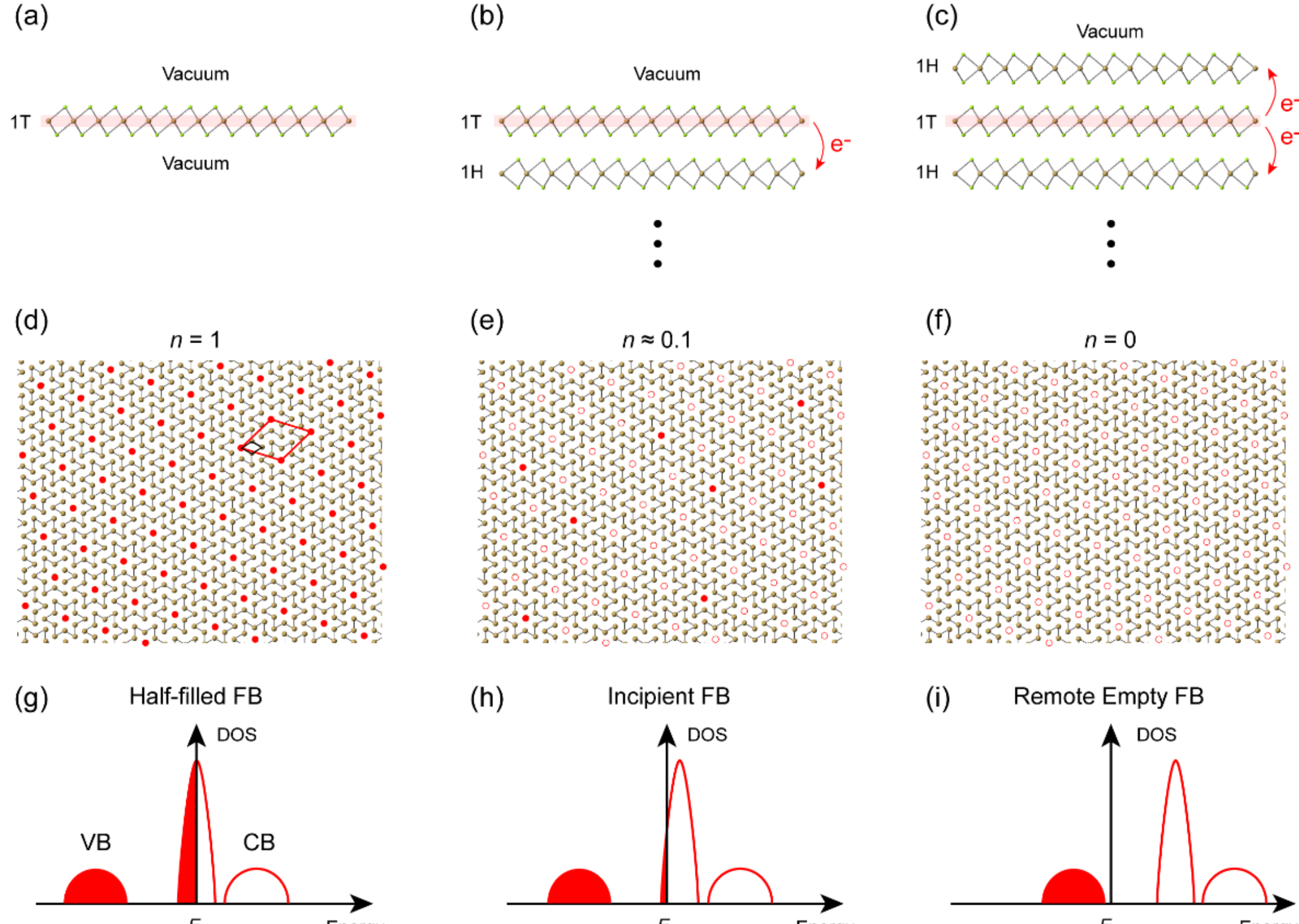

FIG. 2. Hole doping of the 1T-layers. (a)-(c) Schematic illustration of the charge transfer on the free-standing 1T-layer (a), the 1T/1H heterostructure bilayer on top of the 1T-termination (b), and 1H/1T/1H heterostructure tri-layer on top of the 1H-termination (c). (d)-(f) Schematics of the spatially distributed SoDs of the corresponding 1T-layers marked by the red shaded areas in (a)-(c). The electron-filled and void centers of SoDs are marked by red filled circles and hollow circles, respectively. (g)-(i) Band schematics of the corresponding 1T-layers marked by the red shaded areas in (a)-(c). $n$ is the electron-filling factor of the flat band (FB). The original undistorted unit cell and the $\sqrt{13} \times \sqrt{13}$ supercell are indicated by the black and red rhombi in (d), respectively.

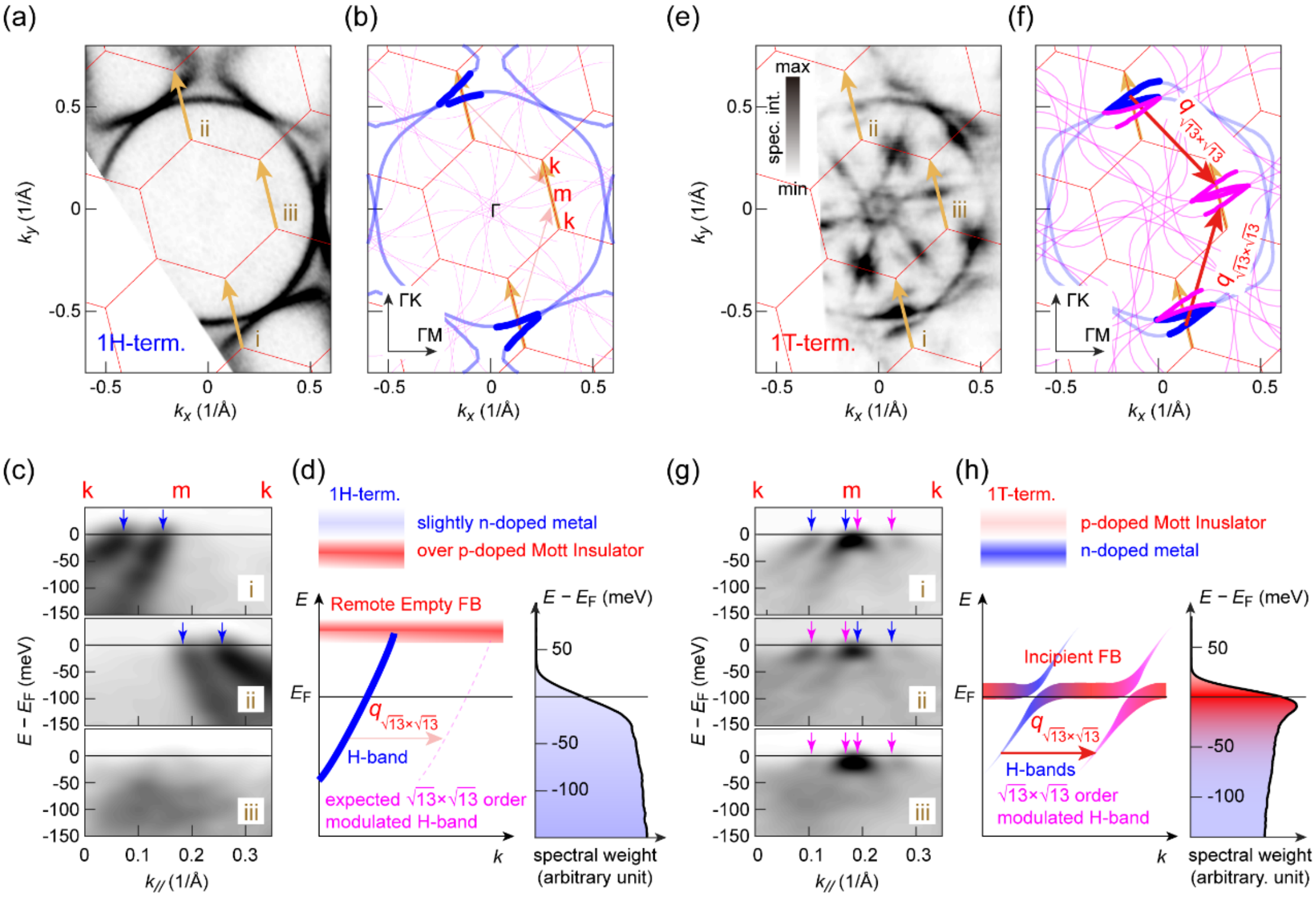

FIG. 3. Interlayer interaction between 1H metallic states and 1T incipient flat bands. (a),(b),(e),(f) ARPES measured (a),(e) and simulated (b),(f) Fermi surfaces on the 1H-termination (a),(b) and 1T-termination (e),(f), appended with the mini-BZ of the $\sqrt{13}\times\sqrt{13}$ superstructure of the 1T-layer (the red hexagon). (c),(g) Band dispersions along k-m-k of the mini-BZ, as indicated by the yellow arrows in (a) and (e), respectively. (d),(h) Top panel: Illustrations of the top two surfaces on the 1H- and 1T-terminations, respectively. Bottom left panel: Schematics of the band structures of the 1H- and 1T-terminations, respectively. Bottom right panel: Integrated energy-distribution-curves (EDCs) measured on the 1H- and 1T-terminations, respectively. Bands derived from 1H-layers, 1T-layers, and scattered states due to 1T/1H interlayer coupling are indicated by blue, red, and magenta curves in (b),(d),(f),(h). The Fermi crossings of the band dispersions derived from 1H-layers, 1T-layers and scattered states due to 1T/1H interlayer coupling are marked by blue, red and magenta arrows in c and g, and highlighted by a thicker linewidth in (b) and (f). The spectral weight is illustrated by different linewidths. The $\sqrt{13}\times\sqrt{13}$ CDW vectors are indicated by red arrows in (b),(d),(f),(h).

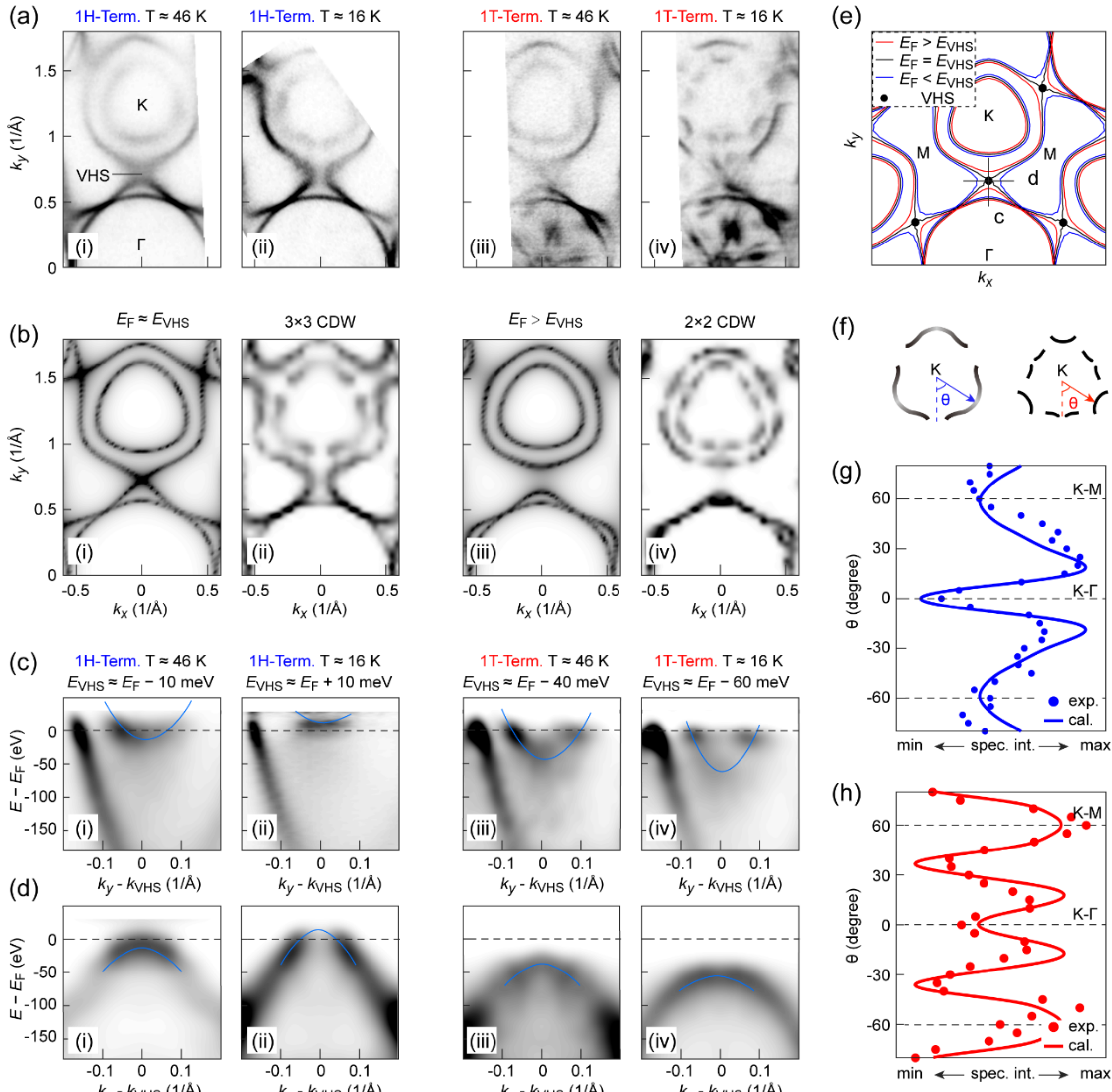

FIG. 4. The VHS and charge ordered electronic states of 1H-layers. (a) Fermi surface mapping measurements of the 1H-termination (i),(ii) and 1T-termination (iii),(iv) in the normal state (i),(iii) and CDW states (ii),(iv) of the surface and subsurface 1H-layers. (b) Corresponding first-principles calculations of (a). (c),(d) Band dispersions across the VHS in (a), measured along $k_y$ (c) and $k_x$ (d) as indicated in (e). The blue curves are guides to the eye of the saddle band near the VHS. (e) Schematic Fermi surfaces of the 1H-layer at $E_F > E_{VHS}$ (red curves), $E_F = E_{VHS}$ (black curves), $E_F < E_{VHS}$ (blue curves). The black dots indicate the momentum positions of the VHS. (f) Schematic Fermi surfaces of the surface 1H-layer (left panel) and the subsurface 1H-layer (right panel). (g),(h) Polar-angle dependent spectral weight distributions of the inner K-centered circular pockets of the surface (g) and subsurface (h) 1H-layers. The polar angles are defined in (f).

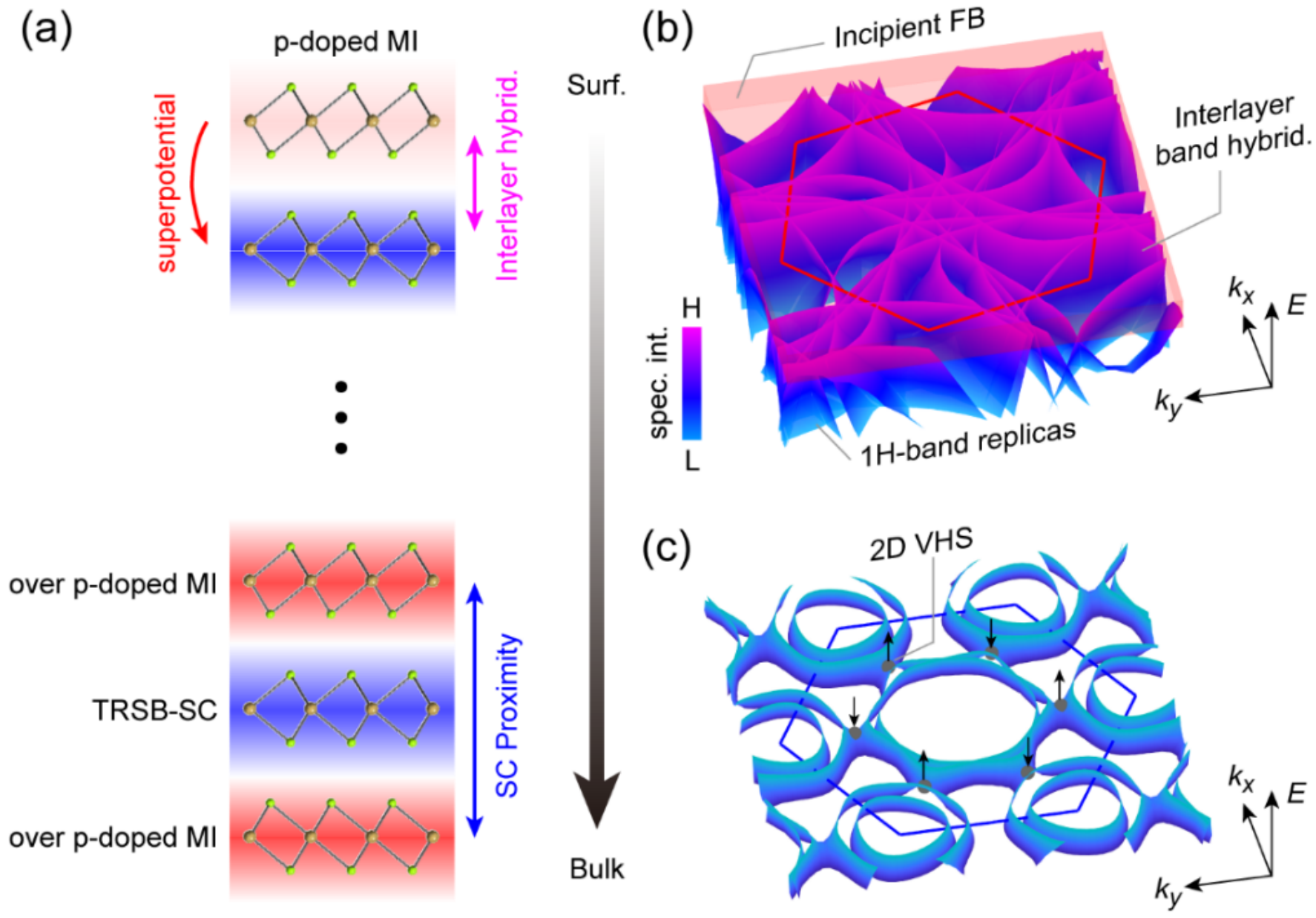

FIG. 5. Layer-dependent electronic structures. (a) Schematics of the layered structure of 4Hb-$TaS_2$: the surface 1T-layer is a heavily p-doped Mott insulator (MI), the bulk 1T-layer an over-doped MI, and the bulk 1H-layer a time-reversal-symmetry-breaking superconductor (TRSB-SC). Umklapp scattering by the $\sqrt{13}\times\sqrt{13}$ superpotential, interlayer band hybridization, and superconducting proximity are marked by red, magenta, and blue arrows. (b) Schematics of the surface Fermi surfaces comprising of interlayer hybridized 1H-band replicas and the incipient FB of the surface 1T-layer. (c) Schematics of the bulk Fermi surfaces with spin-polarized 2D VHS. The spin polarization is denoted by the black arrow. The blue and red hexagons in (c) and (b) are original BZ and the mini-BZ of the $\sqrt{13}\times\sqrt{13}$-superstructure, respectively.